\documentclass[10pt]{article}

 \newcommand{\ben}{\begin{enumerate}}
\newcommand{\een}{\end{enumerate}}
\newcommand{\beq}{\begin{equation}}
\newcommand{\eeq}{\end{equation}}
\newcommand{\bse}{\begin{subequation}}
\newcommand{\ese}{\end{subequation}}
\newcommand{\bea}{\begin{eqnarray}}
\newcommand{\eea}{\end{eqnarray}}
\newcommand{\bc}{\begin{center}}
\newcommand{\ec}{\end{center}}

\def\DR{\rm I\kern-1.45pt\rm R}
\def\DC{\kern2pt {\hbox{\sqi I}}\kern-4.2pt\rm C}
\def\DH{\rm I\kern-1.5pt\rm H\kern-1.5pt\rm I}
\begin{document}

\begin{center}
{\Large \bf Coulomb-Oscillator Duality \\and \\[2mm]
Scattering Problem in 5-Dimensional Coulomb Field}\\
[3mm]
{\large Levon Mardoyan}\\[3mm]
International Center for Advanced Studies, \\Yerevan State
University, Alex Manoogian 1, 375025, Yerevan, Armenia
\end{center}

\vspace{0.5cm}

{\small It is shown that the Hurwitz transformation connects the
eight-dimensional repulsive oscillator problem with the
five-dimensional Coulomb problem for continuous spectrum. The
hyperspherical and parabolic bases for this system are calculated.
The quantum mechanical scattering problem of charged particles in
the 5-dimensional Coulomb field is solved.}

\section {Introduction}

It is known \cite{DMPST,Lambert} that the eight-dimensional
isotropic oscillator \cite{KMT1} with constraint is dual to the
five-dimensional Coulomb problem in the discrete spectrum. This
property is a particular case of a  more general statement which
says that the eight-dimensional isotropic oscillator (without
additional constraint)is dual to the five-dimensional non-Abelian
Yang monopole \cite{MST1,MST2,iwai1}. Similarly, it is possible to
show that the eight-dimensional repulsive oscillator
($U=-\mu\omega^{2}r^{2}/2$) with constraint is dual to the
five-dimensional Coulomb problem in the continuous spectrum. This
fact, just as for the bound states, later was named the
Coulomb-oscillator duality. Due to the $so(5,1)$ hidden symmetry
the Coulomb problem can be factorized not only in hyperspherical
but also in parabolic coordinates. The presence of the parabolic
basis makes possible to construct the scattering theory for the
charged particle in the five-dimensional Coulomb field. But the
dyon-oscillator duality is inherent not only in the  ${\rm I
\!R}^8 \to {\rm I \!R}^5$ mapping but also in the mappings ${\rm I
\!R}^1 \to {\rm I \!R}^1$, ${\rm I \!R}^2 \to {\rm I \!R}^2$, and
${\rm I \!R}^4 \to {\rm I \!R}^3$. As a result, we obtain one- and
two-dimensional anions in the first two cases \cite{ter, ntt} and
the Dirac's monopole in the third \cite{nt1,nt2,MST3}.

This article has the following structure. In section 2 it is shown
that the eight-dimensional repulsive oscillator is dual to the
five-dimensional Coulomb problem for the continuous spectrum.
Section 3 presents the wave functions in five-dimensional
hyperspherical and parabolic coordinates. In section 4 the quantum
mechanical problem of the scattering of the charged particle in
the five-dimensional Coulomb field is considered. The formulae for
the amplitude scattering and cross section are calculated.

\setcounter{equation}{0}

\section {Coulomb-oscillator duality}

Let us consider the Hurwitz transformation \cite{DMPST}:
\begin{eqnarray}
x_0 &=&
u^{2}_{0}+u^{2}_{1}+u^{2}_{2}+u^{2}_{3}-u^{2}_{4}-u^{2}_{5}-u^{2}_{6}-u^{2}_{7},\nonumber
\\ [2mm]
x_{2}+ix_{1} &=&
2\left[\left(u_{0}+iu_{1}\right)\left(u_{5}+iu_{4}\right)+\left(u_{2}-iu_{3}\right)
\left(u_{7}-iu_{6}\right)\right], \\ [2mm] x_4 + ix_3 &=&
2\left[\left(u_{0}+iu_{1}\right)\left(u_{7}+iu_{6}\right)+\left(u_{2}-iu_{3}\right)
\left(u_{5}-iu_{4}\right)\right].\nonumber \label{hurw}
\end{eqnarray}
Here $u_\mu\,(\mu=0,1,...,7)$ are the coordinates of the space
$\rm I\!R^8(\vec{u})$, and $x_i\,(i=0,1,...4)$, of the space $\rm
I\!R^5(\vec{x})$. It is easily seen from (\ref{hurw}) that the
following equality holds:
\begin{eqnarray}
u^4 = \left(u_0^2+u_1^2 + \cdots + u_7^2\right)^2 = x_0^2+x_1^2 +
\cdots + x_4^2 = r^2, \label{euler}
\end{eqnarray}
which is called the Euler identity. According to \cite{DMPST}, the
connection of the Laplace operators in the spaces $\rm I\!R^8$ and
$\rm I\!R^5$ has the form
\begin{eqnarray}
\Delta_8 = 4r\Delta_5 - \frac{4}{r}{\hat J}^2, \label{lap}
\end{eqnarray}
where ${\hat J}^2={\hat J}_1^2+{\hat J}_2^2+{\hat J}_3^2$, and
\begin{eqnarray}
{\hat J}_1 = \frac{i}{2}\left(u_1\frac{\partial}{\partial u_0}-
u_0\frac{\partial}{\partial u_1}+u_3\frac{\partial}{\partial u_2}-
u_2\frac{\partial}{\partial u_3}+u_5\frac{\partial}{\partial u_4}-
u_4\frac{\partial}{\partial u_5}+u_7\frac{\partial}{\partial u_6}-
u_6\frac{\partial}{\partial u_7}\right), \nonumber \\ [3mm] {\hat
J}_2 = \frac{i}{2}\left(u_2\frac{\partial}{\partial u_0}-
u_3\frac{\partial}{\partial u_1}-u_0\frac{\partial}{\partial u_2}+
u_1\frac{\partial}{\partial u_3}-u_6\frac{\partial}{\partial u_4}+
u_7\frac{\partial}{\partial u_5}+u_4\frac{\partial}{\partial u_6}-
u_5\frac{\partial}{\partial u_7}\right), \\ [3mm] {\hat J}_3 =
\frac{i}{2}\left(u_3\frac{\partial}{\partial u_0}+
u_2\frac{\partial}{\partial u_1}-u_1\frac{\partial}{\partial u_2}-
u_0\frac{\partial}{\partial u_3}-u_7\frac{\partial}{\partial u_4}-
u_6\frac{\partial}{\partial u_5}+u_5\frac{\partial}{\partial u_6}+
u_4\frac{\partial}{\partial u_7}\right). \nonumber \label{Ja}
\end{eqnarray}
Using the explicit form of the operators one can prove by a direct
calculation that the operators ${\hat J}_1$, ${\hat J}_2$ and
${\hat J}_3$ satisfy the commutation relations
\begin{eqnarray*}
\left[{\hat J}_a, {\hat J}_b\right] = i\epsilon_{abc} {\hat J}_c,
\label{comJa}
\end{eqnarray*}
where $a, b$ and $c$ are equal 1, 2 and 3, respectively.

Now let us connect the eight-dimensional problem of repulsive
oscillator
\begin{eqnarray}
\left(-\frac{\hbar^2}{2\mu}\Delta_{8} -
\frac{\mu\omega^2u^2}{2}\right)\psi(\vec u) = E\psi(\vec u)
\label{oshr}
\end{eqnarray}
with the five-dimensional Coulomb problem. Since, the operators
$\hat{J}_{a}$ are independent of the coordinates $x_{i}$, we can
represent the wave function $\psi(\vec u)$ of the
eight-dimensional repulsive oscillator  in the following
factorized form
\begin{eqnarray}
\psi(\vec u)=\psi(\vec r)\Phi(\Omega_{a}), \label{os}
\end{eqnarray}
where $\Omega_{a}$ denotes the angles, on which the operators
$\hat{J}_{a}$ depend, and $\Phi(\Omega_{a})$ is the eigenfunction
of the operator $\hat{J}^{2}$, i.e.
\begin{eqnarray}
 \hat{J}^{2}\Phi(\Omega_{a})= J(J+1)\Phi(\Omega_{a}).\label{osJ}
\end{eqnarray}
Here $J(J+1)$ are the eigenvalues of the operator $\hat{J}^{2}$.
Now, substituting (\ref{lap}) into (\ref{oshr}) and taking into
account (\ref{osJ}) we arrive at the equation
\begin{eqnarray}
\left[-\frac{\hbar^2}{2\mu}\Delta_{5} -
\frac{e^2}{r}+\frac{\hbar^{2}}{2\mu r^{2}}J(J+1)\right]\psi(\vec
r) = \varepsilon \psi(\vec r), \label{cshr}
\end{eqnarray}
where $\varepsilon = \mu_0\omega^2/8$ and $4e^2 = E$. Thus, we
obtain that the eight-dimensional repulsive oscillator is dual to
the infinite number of five-dimensional Coulomb systems with
additional terms $1/r^{2}$ and with coupling constant
$\hbar^{2}J(J+1)/2\mu$. At $J=0$ we arrive at the equation for the
ordinary Coulomb problem:
\begin{eqnarray}
\left(-\frac{\hbar^2}{2\mu}\Delta_{5} -
\frac{e^2}{r}\right)\psi(\vec r) = \varepsilon \psi(\vec r),
\label{cshr1}
\end{eqnarray}
The condition $J=0$ is equivalent to requirement
$\hat{J}_{a}\psi(\vec u)=0$. Moreover, it follows from
(\ref{hurw}) that $\psi(\vec x)$ is the even function of variables
$u$
\begin{eqnarray}
\psi\left(\vec x(-\vec u)\right) = \psi\left(\vec x(\vec
u)\right). \label{even}
\end{eqnarray}

\setcounter{equation}{0}

\section {Hyperspherical and parabolic bases}

Let us introduce the five dimensional hyperspherical coordinates
in ${\rm I \!R}^5$ as follows:
\begin{eqnarray}
x_0 &=& r\cos \theta  \nonumber \\ [2mm] x_2 + ix_1 &=& r \sin
\theta \sin \frac{\beta}{2}e^{i\frac{\alpha -\gamma}{2}}  \\ [2mm]
x_4 + ix_3 &=& r \sin \theta \cos \frac{\beta}{2}e^{i\frac{\alpha
+ \gamma}{2}}, \label{sphcoor} \nonumber
\end{eqnarray}
where $r\in [0,\infty)$, $\theta \in [0,\pi]$, $\alpha \in
[0,2\pi)$, $\beta \in [0,\pi]$, $\gamma \in [0,4\pi)$. In this
coordinates the differential elements of length, volume and
Laplace operator have the form
\bea
dl^{2}_{5}&=&dr^{2}+r^{2}d\theta^{2}+\frac{r^{2}}{4}\sin^{2}\theta
\left(d\alpha^{2}+d\beta^{2}+d\gamma^{2}+2\cos\beta d\alpha
d\gamma \right),\\ dV_{5} &=& \frac{r^4}{8} \sin^3 \theta
\sin \beta dr d\theta d\alpha d\beta d\gamma,  \label{svol} \\[2mm] 
\Delta_{5} &=& \frac{1}{r^4}\frac{\partial}{\partial r}
\left(r^4 \frac{\partial}{\partial r}\right) + \frac{1}{r^2 \sin^3
\theta}\frac{\partial}{\partial \theta} \left(\sin^3 \theta
\frac{\partial}{\partial \theta}\right) - \frac{4{\hat L}^2}{r^2
\sin^2\theta}, \label{slap}\eea
where
\begin{eqnarray}
{\hat L}^2 = -\left[\frac{\partial^2}{\partial \beta^2}+ \cot
\beta\frac{\partial}{\partial \beta}+
\frac{1}{\sin^2\beta}\left(\frac{\partial^2}{\partial {\alpha}^2}-
2\cos \beta \frac{\partial^2}{\partial \alpha \partial \gamma}+
\frac{\partial^2}{\partial {\gamma}^2}\right)\right]. \label{Lsq}
\end{eqnarray}
The components of the momentum operator $\hat{\textbf{L}}$ have
the form
\begin{eqnarray}
{\hat L}_1 &=& i\left(\cos{\alpha}\cot{\beta}
\frac{\partial}{\partial \alpha}+
\sin{\alpha}\frac{\partial}{\partial \beta} -
\frac{\cos{\alpha}}{\sin{\beta}} \frac{\partial}{\partial
\gamma}\right) \nonumber \\ [3mm] {\hat L}_2 &=&
-i\left(\sin{\alpha}\cot{\beta} \frac{\partial}{\partial \alpha} -
\cos{\alpha}\frac{\partial}{\partial \beta} -
\frac{\sin{\alpha}}{\sin{\beta}} \frac{\partial}{\partial
\gamma}\right)  \\ [3mm] {\hat L}_3 &=& i\frac{\partial}{\partial
\alpha}, \qquad {\hat L}_3' = i\frac{\partial}{\partial \gamma},
\nonumber \label{La}
\end{eqnarray}

The solution of the equation (\ref{cshr1}) in hyperspherical
coordinates (\ref{sphcoor}) has the following form
\begin{eqnarray}
\psi_{k\lambda Lmm'} \left(r,\theta,\alpha,\beta,\gamma
\right)=\sqrt{\frac{2L+1}{2\pi^{2}}} R_{k \lambda}\left(r\right)
Z_{\lambda L} \left(\theta\right)
D^{L}_{mm'}\left(\alpha,\beta,\gamma\right), \label{swv}
\end{eqnarray}
where Wigner D-function, normalized by the condition
\begin{eqnarray}
\frac{1}{8}\int\left|D^{L}_{mm'}\left(\alpha,\beta,\gamma\right)\right|^{2}
\sin\beta d\alpha d\beta d\gamma = \frac{2\pi^{2}}{2L+1}
\label{ndf}
\end{eqnarray}
is the eigenfunction of the set commuting operators $\hat{L}^{2},
\hat{L}_{3}$ and $\hat{L}_{3'}$ \cite{varsh}. The function
$Z_{\lambda L} \left(\theta\right)$ is given by the formula
\cite{KMT2}
\begin{eqnarray}
Z_{\lambda L} \left(\theta\right) =
2^{2L+1}\Gamma\left(2L+\frac{3}{2}\right)\sqrt{\frac{(2\lambda
+3)(\lambda -2L)!}{2\pi(\lambda
+2L+2)!}}(\sin\theta)^{2L}C_{\lambda
-2L}^{2L+3/2}\left(\cos\theta\right), \label{Zf}
\end{eqnarray}
where $C_{n}^{\lambda}(x)$ are Gegenbauer polynomials, and quantum
number $\lambda$ assumes the values $\lambda = 2L,2L+1,\dots.$

The radial wave function for the continuous spectrum has the form
\begin{eqnarray}
R_{k\lambda}(r) = C_{k\lambda}\frac{(2ikr)^{\lambda}}{(2\lambda +
3)!} {\rm e}^{-ikr} F\left(\lambda + 2 + \frac{i}{ak}; 2\lambda +
4; 2ikr\right), \label{rwv}
\end{eqnarray}
where $k = \sqrt{2\mu\epsilon}/\hbar$, and $a={\hbar}^2/\mu e^2$
is the Bohr radius.

The representation for the confluent hypergeometric function is
\cite{Landau}
\begin{eqnarray}
F\left(a; c; z \right) = \frac{\Gamma(c)}{\Gamma(c-a)}(-z)^{-a}
G\left(a; a-c+1; -z \right) + \frac{\Gamma(c)}{\Gamma(a)} {\rm
e}^z z^{a-c} G\left(c-a; 1-a; z \right), \label{rchf}
\end{eqnarray}
where
\begin{eqnarray}
G\left(a; c; z \right) = 1 + \frac{a c}{1! z} + \frac{a(a+1)
c(c+1)}{2! z^2} + \cdots, \label{G}
\end{eqnarray}
makes it possible to obtain the following expression for $R_{k
\lambda}(r)$:
\begin{eqnarray}
R_{k \lambda}(r) = C_{k \lambda}\frac{(-i)^\lambda}{2k^2r^2} {\rm
e}^{-\pi/2ak} \Re e \left\{\frac{{\rm e}^{-i\left[kr -
\frac{\pi}{2}(\lambda+2) + \frac{1}{ak}\ln{2kr}\right]}}
{\Gamma\left(\lambda+2 -\frac{i}{ak}\right)} G\left(\lambda+2
-\frac{i}{ak}; \frac{i}{ak}-\lambda-1; -2ikr\right) \right\}.
\label{rwv1}
\end{eqnarray}
The normalization constant $C_{k \lambda}$ is
\begin{eqnarray}
C_{k \lambda} = (-i)^\lambda 4k^2{\rm e}^{\pi/2ak}
\left|\Gamma\left(\lambda+2 -\frac{i}{ak}\right)\right|.
\label{nc}
\end{eqnarray}
if the normalization condition for the radial wave function is
\begin{eqnarray}
\int \limits_{0}^{\infty} r^4 R_{k' \lambda}^{*}(r)R_{k
\lambda}(r) dr = 2\pi \delta\left(k - k'\right). \label{ncond}
\end{eqnarray}

If we now let $r \to \infty$ in formula (\ref{rwv1}) and restrict
ourselves to the first term of the expansion, we obtain the
asymptotic expression
\begin{eqnarray}
R_{k \lambda}(r) \approx \frac{2}{r^2} \sin \left[kr
+\frac{1}{ak}\ln{2kr} - \frac{\pi}{2}(\lambda+1) + \delta_\lambda
\right] \label{asymrwv}
\end{eqnarray}
for the radial wave function $R_{k \lambda}(r)$, where
\begin{eqnarray}
\delta_\lambda = \arg \Gamma\left(\lambda+2 -\frac{i}{ak}\right).
\label{phase}
\end{eqnarray}

Now, we define the parabolic coordinates in ${\rm I \!R}^5$ as
\begin{eqnarray}
x_0 = \frac{1}{2}\left(\xi - \eta\right),  \qquad  x_2 + ix_1 =
\sqrt{\xi \eta} \sin \frac{\beta}{2}e^{i\frac{\alpha -\gamma}{2}},
\qquad  x_4 + ix_3 = \sqrt{\xi \eta} \cos
\frac{\beta}{2}e^{i\frac{\alpha + \gamma}{2}},  \label{pcoor}
\end{eqnarray}
where $\xi, \eta \in [0,\infty)$. The differential elements of
length and volume and Laplace operator in the terms of these
coordinates can be written as
\begin{eqnarray}
dl^2_{5} &=& \frac{\xi + \eta}{4}\left(\frac{d\xi^2}{\xi} +
\frac{d\eta^2}{\eta} \right) + \frac{\xi \eta}{4}\left(d\beta^2 +
d\alpha^2 + 2\cos\beta d\alpha d\gamma + d\gamma^2\right) \\ [2mm]
dV_{5} &=& \frac{\xi \eta}{32}(\xi + \eta)\sin\beta d\xi d\eta
d\beta d\alpha d\gamma \\ [2mm] \Delta_{5} &=& \frac{4}{\xi +
\eta}\left[\frac{1}{\xi} \frac{\partial}{\partial \xi}\left(\xi^2
\frac{\partial}{\partial \xi}\right) +
\frac{1}{\eta}\frac{\partial}{\partial \eta}\left(\eta^2
\frac{\partial}{\partial \eta}\right)\right] - \frac{4}{\xi \eta}
{\hat L}^2 \label{plvL}
\end{eqnarray}
Then the equation (\ref{cshr1}) in the parabolic coordinates
(\ref{pcoor}) has the form
\begin{eqnarray}
\left\{\frac{4}{\xi
+\eta}\left[\frac{1}{\xi}\frac{\partial}{\partial
\xi}\left(\xi^{2}\frac{\partial}{\partial
\xi}\right)+\frac{1}{\eta}\frac{\partial}{\partial \eta
}\left(\eta^{2}\frac{\partial}{\partial
\eta}\right)-\frac{4\hat{L}^{2}}{\xi \eta}\right]\right\}\psi
+\frac{2\mu}{\hbar^{2}}\left(\varepsilon +\frac{2e^{2}}{\xi +\eta
}\right)\psi =0. \label{pshcr}
\end{eqnarray}
After the substitution
\begin{eqnarray}
\psi \left(\xi,\eta,\alpha,\beta,\gamma\right) =
\Phi_{1}(\xi)\Phi_2(\eta)D_{mm'}^{L}(\alpha, \beta, \gamma),
\label{pwv}
\end{eqnarray}
the variables in Eq.(\ref{pshcr}) are separated, and we arrive at
the set of differential equations
\begin{eqnarray}
\frac{1}{\xi}\frac{d}{d \xi}\left({\xi}^2 \frac{d\Phi_1}{d
\xi}\right) + \left[\frac{k^{2}}{4}\xi - \frac{L(L+1)}{\xi}
+\frac{\sqrt{\mu}}{2\hbar}\Omega + \frac{1}{2a}\right]\Phi_1 = 0
\nonumber \\ [3mm]
\\
\frac{1}{\eta}\frac{d}{d \eta}\left({\eta}^2 \frac{d\Phi_2}{d
\eta}\right) + \left[\frac{k^{2}}{4}\eta - \frac{L(L+1)}{\eta} -
\frac{\sqrt{\mu}}{2\hbar}\Omega + \frac{1}{2a}\right]\Phi_2 = 0,
\label{set} \nonumber
\end{eqnarray}
where $\Omega$ is the parabolic separation constant.  The function
$\psi \left(\xi,\eta,\alpha,\beta,\gamma\right)$ normalized by the
condition
\begin{eqnarray}
\frac{1}{4}\int \psi_{k' \Omega' L_{1}m_{1}m'_{1}}^* \psi_{k
\Omega L m m'} \xi \eta (\xi+\eta) d\xi d\eta = 2\pi \delta\left(k
- k'\right) \delta\left(\Omega -
\Omega'\right)\delta_{LL_{1}}\delta_{mm_{1}}\delta_{m'm'_{1}}
\label{pncond}
\end{eqnarray}
leads to the parabolic basis
\begin{eqnarray}
\psi_{k\Omega Lmm'}\left(\xi,\eta,\alpha,\beta,\gamma\right)
=\sqrt{\frac{2L+1}{2\pi^{2}}} '_{k \Omega L} \Phi_{k\Omega
L}(\xi)\Phi_{k -\Omega L}(\eta) D_{mm'}^{L}(\alpha,\beta,\gamma),
\label{parwv}
\end{eqnarray}
where
\begin{eqnarray}
\Phi_{k \Omega L}(x) = \frac{(ikx)^L}{(2L+1)!} {\rm e}^{-ikx/2}
F\left(L+1+\frac{i}{2ak}+\frac{i\sqrt{\mu}}{2\hbar k}\Omega; 2L+2;
ikx\right), \\ [2mm] C_{k \Omega  L} =
(-1)^{L}\sqrt{\frac{\hbar^{2} k^3}{2\pi \mu}}\, {\rm
e}^{\frac{\pi}{2ak}} \left|\Gamma\left(L+1-\frac{i}{2ak}-
\frac{i\sqrt{\mu}}{2\hbar k}\Omega\right)
\Gamma\left(L+1-\frac{i}{2ak}+\frac{i\sqrt{\mu}}{2\hbar k}\Omega
\right)\right|.
\end{eqnarray}
We calculate the normalization constant $C_{k \Omega  L}$ using
the representation (\ref{rchf}).

\setcounter{equation}{0}

\section {The 5-dimensional generalization of the Rutherford's formula}

We now consider the scattering problem of a charged particle in
the five-dimensional Coulomb field. Since the motion in Coulomb
field of arbitrary dimensions $d\geq 3$ is two-dimensional
problem, the wave function is independent of the angles
$\alpha,\,\beta$ and $\gamma$, i.e. independent of the quantum
numbers $L,\,m$ and $m'$. Substituting $L=0$ in Eqs. (\ref{set}),
we obtain
\begin{eqnarray}
\frac{1}{\xi}\frac{d}{d \xi}\left({\xi}^2 \frac{d\Phi_1}{d
\xi}\right) + \left[\frac{k^{2}}{4}\xi
+\frac{\sqrt{\mu}}{2\hbar}\Omega + \frac{1}{2a}\right]\Phi_1 = 0
\nonumber \\ [3mm]
\\
\frac{1}{\eta}\frac{d}{d \eta}\left({\eta}^2 \frac{d\Phi_2}{d
\eta}\right) + \left[\frac{k^{2}}{4}\eta -
\frac{\sqrt{\mu}}{2\hbar}\Omega + \frac{1}{2a}\right]\Phi_2 = 0,
\label{set1} \nonumber
\end{eqnarray}

We seek such solutions of Eqs. (\ref{set1}) that the solution of
the Schr\"{o}dinger equation for negative $x_{0}\in (-\infty; 0)$
and large $r \to \infty$ has the form of a flat wave:
\begin{eqnarray}
\psi_{k\Omega}(\xi, \eta) \sim e^{ikx_{0}} = e^{\frac{ik}{2}(\xi -
\eta)}. \label{planew}
\end{eqnarray}
This condition can be satisfied if we set the parabolic separation
constant equal to
\begin{eqnarray}
\Omega =-\frac{\hbar}{a\sqrt \mu} -i\frac{2\hbar k}{\sqrt \mu}.
\label{omega}
\end{eqnarray}
Substituting this relation in Eqs. (\ref{set1}), we obtain a
solution of the Schr\"{o}dinger equation that describes the
scattering of a charged particle in the five-dimensional Coulomb
field:
\begin{eqnarray}
\psi_{k}(\xi, \eta) = '_ke^{\frac{ik}{2}(\xi - \eta)}
F\left(\frac{i}{ak}; 2; ik\eta\right), \label{psi}
\end{eqnarray}
where $C_k$ is the normalization constant. To separate the
incident and scattered waves in function (\ref{psi}), we must
investigate the behavior of this function at large distances from
the scattering center. Using the first two terms in representation
(\ref{rchf}) for a confluent hypergeometric function, we obtain
\begin{eqnarray}
F\left(\frac{i}{ak}; 2; ik\eta\right) \approx e^{-\frac{\pi}{2ak}}
\left[\frac{e^{-\frac{i}{ak}\ln{k\eta}}}{\Gamma\left(2-\frac{i}{ak}\right)}
\left(1+\frac{1+iak}{ia^2k^3\eta}-\frac{1+a^{2}k^{2}}{2a^{4}k^{6}\eta^{2}}\right)
- \frac{i(ak+i)}{\Gamma\left(2+\frac{i}{ak}\right)}
\frac{e^{ik\eta}}{a^{2}k^{4}\eta^{2}}e^{\frac{i}{ak}\ln{k\eta}}\right].
\label{ef11}
\end{eqnarray}
for the large $\eta$. We now substitute this relation in wave
function \ref{psi}) and select the normalization constant $C_k$ in
the form
\begin{eqnarray}
C_k = e^{\pi/2ak}\Gamma\left(2- \frac{i}{ak}\right) \label{cscat}
\end{eqnarray}
for the incident wave to have a unit amplitude. Using the formulae
$r = (\xi + \eta)/2$ and $\eta = r - x_{0} = r(1-\cos \theta)$ to
change to spherical coordinates, we obtain
\begin{eqnarray}
\psi_{k}(\xi, \eta) =
\left[1+\frac{ak-i}{2a^2k^3r\sin^2\theta/2}\right]
\exp\left[ikx_{0}-
\frac{i}{ak}\ln{\left(2kr\sin^2\frac{\theta}{2}\right)}\right] +
\\ [3mm]
+ \frac{f(\theta)}{r^{2}} \exp\left[ikr +
\frac{i}{ak}\ln{(2kr)}\right], \label{psiscat}
\end{eqnarray}
where $f(\theta)$ is the scattering amplitude,
\begin{eqnarray}
f(\theta) = \frac{(1-iak)}{4a^{2}k^4\sin^2\theta/2}
\frac{\Gamma\left(2-i/ak\right)}{\Gamma\left(2+i/ak\right)}
\exp\left(\frac{2i}{ak}\ln{\sin \frac{\theta}{2}}\right).
\label{amplit}
\end{eqnarray}

Therefore, for the scattering cross section $d\sigma =
\left|f(\theta)\right|^2d\Omega$ ($d\Omega$ is the element of the
solid angle), we obtain the formula
\begin{eqnarray}
d\sigma = \frac{1+a^2k^2}{16a^4k^8\sin^8\theta/2}d\Omega,
\label{section}
\end{eqnarray}
which generalizes the Rutherford's formula in five-dimensional
case.

\vspace{5mm}

{\large Acknowledgements.} The work is carried out with the
support of ANSEF No: PS81 grant.

\end{document}